\documentstyle[epsf,preprint,aps]{revtex} 
\begin{document}
\draft
 

\title
{Exact Ground States and Excited States of Net Spin Models}
 
\author{ H. Q. Lin$^{1}$ and J. L. Shen$^{1,2}$ }
\address{
$^{1}$Department of Physics, Chinese University of Hong Kong, Hong Kong\\
$^{2}$Institute of Physics, Chinese Academy of Science, Beijing, China
}
 
\date{\today}

\maketitle

\begin{abstract}
We study a set of exactly soluble net spin models.
There exist two kinds of ground state,
one is a complete dimerized state,
and the other one is the ground state of corresponding spin-1 model.
For the excitation gap, various phases were discovered and determined.
\end{abstract}


\newcommand{\beq}{\begin{equation}}
\newcommand{\eeq}{\end{equation}}
\newcommand{\bsigma}{\mbox{\boldmath $\sigma$}}

It is well known that a spin-1/2 antiferromagnetic chain with nearest-neighbor
coupling $J_1$ and next neighbor coupling $J_2$ could dimerize in the presence
of frustration, as exemplified by the exactly soluble one-dimensional
Majumdar-Ghosh model \cite{mg_model},
where the exact (twofold degenerate) ground
state is a simple product of singlet dimers. The elementary excitation can be
constructed as a pair of unbound spins above the completely dimerized state
\cite{shastry-sutherland}. This model also can be considered as a two-chain
with one diagonal or a zigzag spin ladder. Analytical and numerical studies
\cite{haldane,affleck,okamoto} of the model show a transition from a gapless
phase when $J_2 < J_{2c}$ to a gapped phase when $J_2 > J_{2c}$.
Recently, the Majumdar-Ghosh model was studied
by the density matrix renormalization group (DMRG) approach.
The DMRG study of Chitra et al. \cite{chitra} considered a more general model
which included a dimerization term.
White and Affleck \cite{white-affleck} studied the zigzag spin ladder
by using the bosonization and the DMRG methods.
In particular, they pointed out a potential important connection between
the zigzag chain and the generalized Kondo lattice.
In addition, a family of spin-ladder models were studied by Kolezhuk and
Mikeska \cite{kolezhuk-mikeska} that exhibit non-Haldane spin-liquid
properties as predicted by Nersesyan and Tsvelik \cite{nersesyan-tsvelik}.

In this Letter we consider a set of spin-1/2 antiferromagnetic Heisenberg
models defined on a double layer as shown in Fig. 1.
Each layer has $M \times L$ sites and it connects to the other one
not only perpendicularly, but also diagonally.
$L$ is a measure of the dimensionality which could be any integer
varying from 1 (the net spin ladder) to $M$ (the net spin layers).
Throughout this work, we use $\bf S$ to represent spins on the lower layer,
and $\bf S'$ on the upper layer.
Open boundary conditions are imposed.

Under certain combination of coupling constants, we show that for general $L$
and $M$ the model could be solved exactly.
First, 
the model have two kinds of ground state, one is a completely
dimerized state, and the other one is the ground state of the spin-1 model
defined on the single layer.
Second, the model exhibits rich excitation phases, depending on coupling
constants and dimensionality $L$.
As we shall show below, these properties are quite different from the
existing models such as the Majumdar-Ghosh model \cite{mg_model}.

The model Hamiltonian is:
\begin{eqnarray}
H &=& 2J_1 \sum_{k,l=1}^{M,L}   {\bf S}_{k,l}  \cdot {\bf S'}_{k,l}
\nonumber \\
  &+& 2J_2 \hspace{-0.3cm} \sum_{k,l=1}^{M-1,L-1} \hspace{-0.3cm}
        {\bf S}_{k,l}  \cdot ( {\bf S}_{k+1,l}  + {\bf S}_{k,l+1} )
      + {\bf S'}_{k,l} \cdot ( {\bf S'}_{k+1,l} + {\bf S'}_{k,l+1} )
        \nonumber \\
  &+& 2J_3 \sum_{k,l=1}^{M-1,L-1}
        {\bf S'}_{k,l} \cdot ( {\bf S}_{k+1,l}  + {\bf S}_{k,l+1} ) \nonumber\\
  &+& 2J_4 \sum_{k,l=1}^{M-1,L-1}
        {\bf S }_{k,l} \cdot ( {\bf S'}_{k+1,l} + {\bf S'}_{k,l+1} ) ~,
\label{model_h}
\end{eqnarray}
with $J_1, J_2, J_3, J_4 \geq 0$, and $|{\bf S}| = |{\bf S'}| = S = 1/2$.
Both $J_1$ and $J_2$ favor local antiferromagnetic ordering 
while $J_3$ and $J_4$ represent frustration effects.


For any integers $M$ and $L$, we found that a complete dimerized state:
\beq
   \psi_D = [1,1'][2,2'] \cdots [M,M'] \cdots [N,N'] ,~  N=ML ~,
\label{dimer_conf}
\eeq
where $[i,j]$ denotes the normalized singlet
${1 \over \sqrt{2}} ( \alpha_i \beta_j - \beta_i \alpha_j )$
with $\alpha$ and $\beta$ representing the usual up and down single spin
eigenfunctions, respectively,
is an eigenstate of Eq. (\ref{model_h})
when we impose the following constraint:
\beq
   2 J_2 = J_3 + J_4 ~.
\label{dimer_cond}
\eeq
To prove our statement,
we first show that this is true for the 4-spin plaque case,
$\{ {\bf S}_{1,1}, {\bf S'}_{1',1'}, {\bf S}_{2,1'}, {\bf S'}_{2',1'} \}$,
simplied as
$\{ {\bf S}_1, {\bf S'}_{1'}, {\bf S}_2, {\bf S'}_{2'} \}$,
which is indicated in Fig. 1.
The model Hamiltonian commutes with total spin $S_{tot}$
so one can use it to characterize eigenstates.
In the sub-space of zero $z$-component magnetization there are six eigenstates
and among these six eigenstates there are two singlets which are determined by
the following equations:
\begin{eqnarray}
   H \psi_D &=& -3J_1 \psi_D 
	+ {\sqrt{3} \over 2} (2J_2 - J_3 - J_4) \psi_2 \\ 
   H \psi_2 &=& {\sqrt{3} \over 2} (2J_2 - J_3 - J_4) \psi_D
	+ (J_1 - 2J_2 - J_3 - J_4) \psi_2 \nonumber
\label{4-spin}
\end{eqnarray}
where
\begin{eqnarray}
   \psi_D &=& {1 \over 2} ( \alpha_1 \beta_{1'} - \beta_1 \alpha_{1'} )
			( \alpha_2 \beta_{2'} - \beta_2 \alpha_{2'} )
\end{eqnarray}
is the completely dimerized state it becomes an eigenstate of $H$ 
with eigenvalue $-3J_1$ and another singlet state $\psi_2$ with eigenvalue
$E_2 = J_1 - 4J_2$ when Eq. (\ref{dimer_cond}) holds.

Also, we obtained other four eigenvalues according to their total spin
$S_{tot}$, among them there is a triplet with eigenvalue:
\beq
   E_3     = -(J_1 + J_2) + {1 \over 2} (J_3 + J_4) ~. \label{e3}
\eeq

We know that the formation of dimers $[1,1']$ and $[2,2']$ is due to
antiferromagnetic coupling $J_1$.
Without antiferromagnetic couplings $J_3$ and $J_4$, quantum
fluctuation due to $J_2$ will kill both dimers.
As long as condition Eq. (\ref{dimer_cond}) holds, the effect of $J_2$ on
the dimers will be canceled exactly by $J_3$ and $J_4$.
This is also true no matter how many plaques we put together
and thus $\psi_D$ is the eigenstate of Hamiltonian (\ref{model_h})
with eigenvalue $E_D = - {3 \over 2} M J_1$.

It is also obvious that the dimerized state $\psi_D$ may not be the ground
state for general coupling parameters $\{J_1,J_2,J_3,J_4\}$
due to the fact that the antiferromagnetic couplings $J_2, J_3, J_4$ which
connect one plaque to another also introduce interactions between plaques.
Since in the limit of $J_1 \rightarrow \infty$, $\psi_D$ must be the ground
state, and in the limit of $J_1 \rightarrow 0$, $\psi_D$ cannot be the ground
state, therefore we expect that there exists a critical value of $J_{1c}$
such that when $J_1 > J_{1c}$ the ground state is completely dimerized.

We should mention that for the case of $L=1$, the spin net model (we call it
net spin ladder) goes back to
the Majumdar-Ghosh model \cite{mg_model}
when $J_2 = 0.5 J_1, J_3 = J_1$, and $J_4 = 0$.
Several studies on the generalization of the M-G model,
such as the zigzag model
\cite{shastry-sutherland,haldane,affleck,okamoto,chitra,white-affleck},
were carried out before.

To proceed, we study the case of $L=1$ with coupling constants $J_3=J_4=J_2$
in details as an illustration.
This special case has been recently studied by several groups
\cite{bose-gayen,xian,su,whzheng-et,fei,xqwang}.
We use it as an example to present our approach
which leads us to solve the whole class of net spin models.

We can rewrite the Hamiltonian as
\beq
H = - 2M J_1 {3 \over 4} + J_1 \sum_{k=1}^{M-1} {\bsigma}_{k}^2
  +  2J_2 \sum_{k=1}^{M-1} {\bsigma}_{k} \cdot {\bsigma}_{k+1} ~,
\label{model_sone}
\eeq
where ${\bsigma}_{k} = {\bf S}_{k}   + {\bf S'}_{k}$.
This Hamiltonian describes a chain of $M$ spins with either
$|{\bsigma}_{k}| = 0$(singlet) or $1$(triplet).

With this form of $H$,
it is easy to see that $\psi_D$ is an eigenstate of $H$ by using the relation
$( {\bf S}_{k} + {\bf S'}_{k} )[k, k'] = 0$, hence
\beq
 \left( J_1 \sum_{k=1}^{M-1} {\bsigma}_{k}^2
      + 2J_2 \sum_{k=1}^{M-1} {\bsigma}_{k} \cdot {\bsigma}_{k+1}
 \right) |\psi_D> = 0 ~,
\eeq
and the eigenvalue is $E_D = -{3 \over 2} MJ_1$.

To lower energy, the squared term, $J_1 {\bsigma}_{k}^2$, favors to be in
the $|{\bsigma}_{k}| = 0$ state,
while the exchange term, $J_2 {\bsigma}_{k} \cdot {\bsigma}_{k+1}$,
favors to be in the $|{\bsigma}_{k}| = 1$ state.
If all pairs ${\bf S}_{k} + {\bf S'}_{k}$ are in the $\sigma = 1$ state then
the model is equivalent to the spin-1 chain with coupling constant $2J_2$,
apart from a constant ${MJ_1 \over 2}$.
One can see from Eq. (\ref{model_sone}) that the eigenstates of the spin-1
chain could also be the eigenstate of the net spin ladder model.
In fact, as $J_1$ decreases further, the ground state of the net spin ladder
model becomes the ground state of the spin-1 chain.

Moreover, we can use this criterion to determine the transition point $J_{1c}$.
When $J_1 = J_{1c}$, $<H> = E_D = - {3 \over 2} M J_1$,
where the expectation value is taken in the ground state. We obtain
\beq
   < J_{1c} \sum_{k=1}^{M-1} {\bsigma}_{k}^2 
   + 2J_2 \sum_{k=1}^{M-1} {\bsigma}_{k} \cdot {\bsigma}_{k+1} > = 0 ~,
\eeq
i.e., $2J_{1c} M = $ ground state energy of $M$-site spin-1 chain with coupling
constant $2J_2$. Using the existing estimates \cite{white-huse}, we obtain
\beq
	J_{1c} = 1.4015 J_2 ~.
\eeq
To confirm our analysis, we determine the critical value $J_{1c}$ by using
the exact diagonalization technique \cite{hql} for several clusters.
Our numerical answer is
$J_{1c} = 1.402 \pm 0.001$, the same as the above within numerical error.

We next study excitation spectrum of the model. One can create a single magnon 
simply by breaking a dimer:
${1 \over \sqrt{2}} (\alpha_i\beta_j - \beta_i\alpha_j)
\rightarrow \alpha_i \alpha_j$
(or $\beta_i\beta_j$
or ${1 \over \sqrt{2}} (\alpha_i\beta_j + \beta_i\alpha_j)$.)
It can be shown that state
\beq
   \psi_m = [1,1'][2,2'] \cdots (\uparrow, \uparrow)\cdots [M,M'] ~,
\eeq
is also an eigenstate of the net spin ladder model with total spin $S_{tot}=1$.
The energy gap in this case is exactly $\Delta_{st} = 2J_1$ and
the transition is from singlet to triplet.
This excitation energy is independent of other antiferromagnetic couplings
as long as the condition (\ref{dimer_cond}) holds
and thus it is possible that there exists states with lower excitation energy.

Analyzing results for the single plaque, we found that
when $J_1 < 2 J_2$, $E_2 = J_1 - 4J_2$ is lower than
$E_3 = - J_1$ of Eq. (\ref{e3}).
The former is in the singlet ($S_{tot}=0$) sector.
Interesting enough, this critical value $(J_1/J_2)_c = 2$ is independent of
lattice size.
This could be easily understood by using the Hamiltonian of
Eq. (\ref{model_sone}).
Such excitation corresponds to the situation
where two adjunct dimers are broken (so they are in the spin $S=1$ state)
while the rest of dimers remain unbroken, regardless the chain length.
The two spin 1 sites form a singlet with energy $-4J_2$
so the excitation energy, which corresponds to the transition from
the dimer singlet to another singlet, is $\Delta_{ss} = 4J_1 - 4J_2$.
$\Delta_{ss} < \Delta_{st}$ when $J_1 < 2J_2$.

What will happen as one decreases $J_1$ further? 
We found that the first excited state remains to be  the one we just described
for the whole region of $2J_2 > J_1 > J_{1c}$.
Moreover, this state is $(M-1)$-fold degenerate.
States consisting of longer segment of broken dimers are high in energy.

Figure 2a summarizes our findings.
Here gap $\Delta$ is defined as the difference
between the first excited state energy and the ground state energy,
$E(1) - E(0)$.
There are three regions, depending on parameter $J_2/J_1$:

\begin{description}
\item[Region I], $J_1 \geq 2 J_2$,
$\Delta/J_1=2$, independent of $J_2$ and it is singlet to triplet.

\item[Region II], $2 J_2 > J_1 > J_{1c}$,
$\Delta/J_1 = 4 ( 1 - J_2/J_1 )$ decreases linearly from $2$ to 
$4 ( 1 - J_2/J_{1c} ) = 1.1459$ at $J_1 = J_{1c} +$.
The excitation is from singlet (dimers) to another singlet.
In both regions I and II, the completely dimerized state is the ground state.

However, right at $J_1 = J_{1c}$, the completely dimerized state
is degenerate with the ground state of spin-1 chain.

\item[Region III], $J_1 > J_{1c}$, it is the Haldane gap phase.
At $J_1 = J_{1c} -$, $\Delta/J_1 = 0.5851$.
\end{description}

%


We now turn to the double layer case ($L=M$ and $N=LM$).
Similarly, we get
\begin{eqnarray}
H &=& - 2N J_1 {3 \over 4} + J_1 \sum_{k=1}^N {\bsigma}_{k}^2
   + 2J_2 \sum_{<k,l>}^N {\bsigma}_{k} \cdot {\bsigma}_{l} ~,
\label{model_s12d}
\end{eqnarray}
where ${\bsigma}_{k} = {\bf S}_{k,l}   + {\bf S'}_{k,l}$
and $<k,l>$ refer to nearest neighbors in two dimensions.
Again, when all $|{\bsigma}| = 1$, this is nothing but the two-dimensional
antiferromagnetic Heisenberg model with coupling constant $2J_2$,
apart from a constant.

With the same approach we had described above, it is easy to see that a
complete dimerized state where all $\bsigma$ are in singlet state is the
eigenstate of the model and it is the ground state when $J_1$ is sufficiently
large. The critical point at which the complete dimerized state becomes the
excited state is determined by,
\beq
J_{1c} = - {J_2 \over N} \sum_{<k,l>}^N {\bsigma}_{k} \cdot {\bsigma}_{l} ~.
\eeq
Using the existing estimates \cite{hql-emery}, we obtain
\beq
J_{1c} = 2.3323 J_2 ~.
\eeq

Two interesting observations are right in hand:

(1) The existence of the critical point is similar to that for the double
layer layer antiferromagnetic Heisenberg model without diagonal coupling
($J_3$ and $J_4$ set to be zero) terms.
There has been a lot of studies on that model, by the spin-wave theory,
perturbation theory, bond-operator approach, etc. (Ref. [19] and references
therein.)
The transition point at which gap vanishes is $J_1/J_2 = 2.54$,
close to the value of $J_{1c}$ we got here, although with different model.

(2) We can show that it is still true when $J_1/J_2 \leq 2$, the first
excitation is a triplet. 
Moreover, after $J_1/J_2 \leq 2.3323$, the system is gapless because the
existence of long-range-order in the two-dimensional spin-1 Heisenberg model,
as shown rigorously long time ago \cite{dyson-et,neves-perez}.
Therefore, there exists no transition from the dimer singlet
to another singlet as what we have seen for the net spin ladder case.
Note that the numerical accuracy of $2.3323$ is unimportant here,
as long as the ground state energy per site is lower than $-2J_2$,
(which must be true because $-2J_2$ just corresponds to the Neel state!),
our assertion holds.
We plot the excitation gap as function of $J_2/J_1$ in Fig. 2b.

What will happen for cases of $1 < L < \infty$?
We have studied two cases, $L=2$ and $L=3$.
For $L=2$, the critical value $J_{1c}$ is determined by the corresponding 
spin-1 model defined on the $2 \times M$ lattice.
Table I. lists the ground state energy per site as well as per bond.
Total number of lattice sites is $N = 2 \times M$ and total number of bonds
is $N_b = 3 \times M - 2$ when open boundary is used.


\noindent It is clear that, in the limit of $N \rightarrow \infty$, we have
\begin{eqnarray}
E/NJ_2 &=& (E/N_b) (N_b/N) = (3/2) (E/N_b) \\ \nonumber
    &>& (1.5) * (-1.3) = -1.95 > -2.0.
\end{eqnarray}
Thus, $J_{1c} < 2 J_2$ and the excitation spectrum for the case of $L=2$
is similar to that of $L=1$.

Similarly, we list the ground state energy per site as well as per bond in
Table II for the case of $L=3$.


\noindent Because we do not have enough data due to lack of computer power,
we cannot confidently establish a lower bound for the ground state energy
in the thermodynamics limit.
Although a second order perturbation theory calculation gives
\beq E/NJ_2 = - {617 \over 315} > {630 \over 315} = -2 ~, \eeq
it is too close to be definitive.
On the other hand, the second order perturbation theory calculation for
the $L=2$ case gives
\beq E/NJ_2 = -{9 \over 5} > -2 ~,\eeq
which is more definitive.

Therefore, we cannot rule out a possibility of dimensional crossover from 1-D
to 2-D occurs at $L=3$.
It is also equal probable that $J_{1c} < 2 J_2$ but the difference between
$J_{1c}$ and $2J_2$ becomes exponentially small when $L$ increases from 3.

For other combinations of couplings, one can only determine the critical points
and eigenvalues of the model numerically.
However, qualitative behaviors should be the same as what we have shown here.
Note that the energy gap $2J_1$ is quite large (critical values of $J_2$
are smaller than $J_1$) so the phase diagram should be similar to Fig. 2.
Also we note that one obtains a critical value of $J_1/J_2 = 2.54$ when
$J_3 = J_4 = 0$ is very close to that when $J_3 = J_4 = J_2 = (1/2.33)J_1$.
This shows that our results are insensitive to the values of $J_3$ and $J_4$.
Essential physics is determined by the two couplings: $J_2$ and $J_1$.

In summary,
we have studied a class of net spin models,
defined by the Hamiltonian Eq. (\ref{model_h}),
and controlled by three antiferromagnetic couplings:
$J_2/J_1$, $J_3/J_1$, $J_4/J_1$, and dimensionality $L$.
When $2 J_2 =  J_3 + J_4 $,
the completely dimerized state $\psi_D$ is an eigenstate of these models
and it is also the ground state if $J_1$ is sufficiently large.
To fully understand the model,
we have studied the case of $J_2 = J_3 = J_4$ extensively
and obtained phase diagrams for both 1-D(ladder, $L=1$)
and 2-D(double layer, $L=M$) cases.
For the ladder case at a particular point in the parameter space:
$J_2 = {1 \over 2} J_1$, $J_3 = J_1$, and $J_4 = 0$
this model returns to the Majumdar-Ghosh model \cite{mg_model}.
But its behavior is quite different from that of the Majumdar-Ghosh model.
We also show qualitative different behavior between the ladder and the double
layer cases.
Moreover, we have studied the cases of $L=2$ and $L=3$ and discussed
the possibility of crossover from 1-D to 2-D.
In addition, we made connection with other studies of double layer system.

We are grateful to Dr. Kolezhuk for valuable e-mail communication.
We thank Dr. Yun Song, Mr. Qiang Gu, and Mr. Hon Lung Mak
for technical assistance.
This work was supported in part by the Earmarked Grant for Research from the
Research Grants Council (RGC) of the Hong Kong Government
under project CUHK 311/96P. 


\newpage
\begin{figure}[m]
\hspace*{-1.0cm}
\epsfysize=5.5in\epsfbox{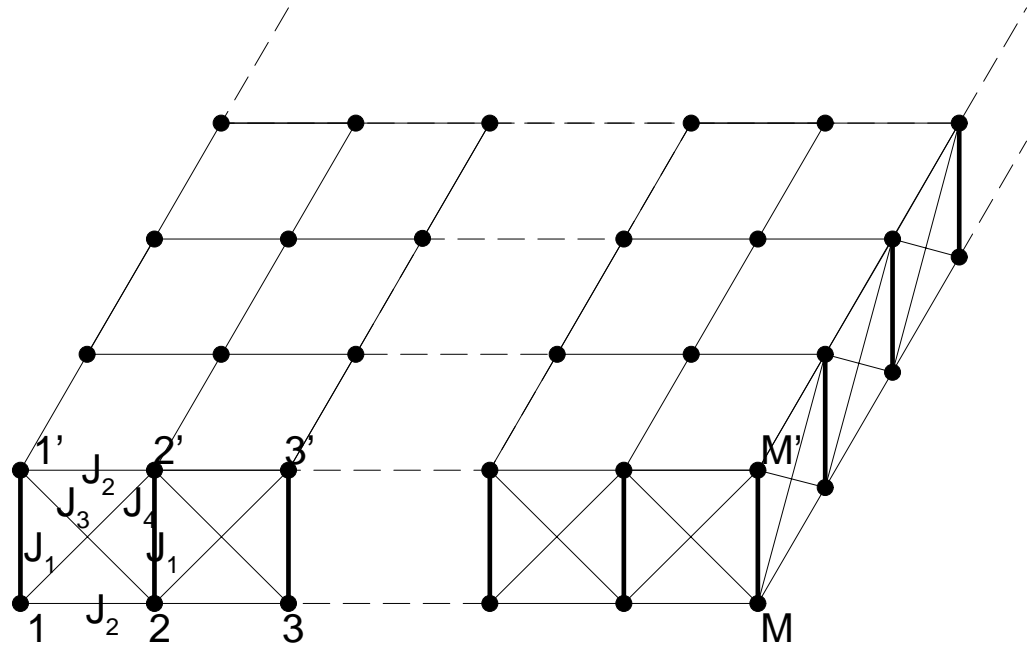}
\vskip 1.0cm
\caption{The net spin layers of $2ML$ spins,
where thick solid lines represent singlet dimers.}
\label{fig:layer}
\end{figure}

\newpage
\begin{figure}[m]
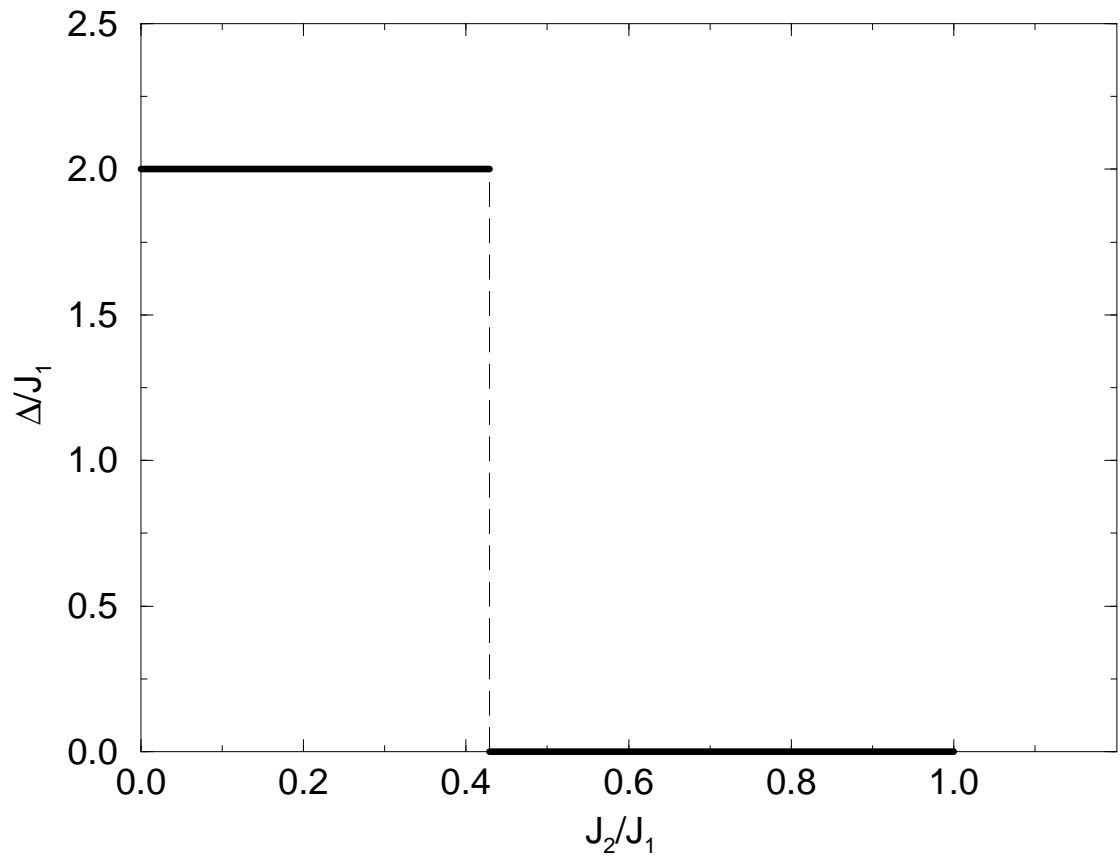

\centerline{\large Fig. 2a}
\vspace{1.0cm}
\epsfysize=5.0in\epsfbox{netspin_fig2a}
\newpage
\centerline{\large Fig. 2b}
\vspace{1.0cm}
\epsfysize=5.0in\epsfbox{netspin_fig2b}
\caption{Excitation gap as function of coupling parameters for
the net spin ladder (a), and double layer (b) models.}
\label{fig:phase}
\end{figure}

\newpage
\hspace*{2.0cm} Table I
\vskip 1.0truecm
\begin{tabular}{|c|l|l|} \hline\hline
Lattice      &
   \hspace{0.50cm}$E(0) / N $ & \hspace{0.25cm} $E(0) / N_b$ \\ \hline
$2 \times 2$ &  -1.50000000 & -1.50000000 \\ \hline
$2 \times 3$ &  -1.61633441 & -1.38542949 \\ \hline
$2 \times 4$ &  -1.68229677 & -1.34583741 \\ \hline
$2 \times 5$ &  -1.72097660 & -1.32382816 \\ \hline
$2 \times 6$ &  -1.74706131 & -1.31029598 \\ \hline
$2 \times 7$ &  -1.76571007 & -1.30104952 \\ \hline
$2 \times 8$ &  -1.77973127 & -1.29435001 \\ \hline
\end{tabular}
\vskip 0.5truecm

\vskip 2.0truecm
\hspace*{2.0cm} Table II
\vskip 1.0truecm
\begin{tabular}{|c|l|l|} \hline\hline
Lattice      &
   \hspace{0.50cm}$E(0) / N $ & \hspace{0.25cm} $E(0) / N_b$ \\ \hline
$3 \times 2$  & -1.61633441 & -1.38542949 \\ \hline
$3 \times 3$  & -1.71359964 & -1.28519973 \\ \hline
$3 \times 4$  & -1.80819446 & -1.27637256 \\ \hline
$3 \times 5$  & -1.83887667 & -1.25377955 \\ \hline
\end{tabular}

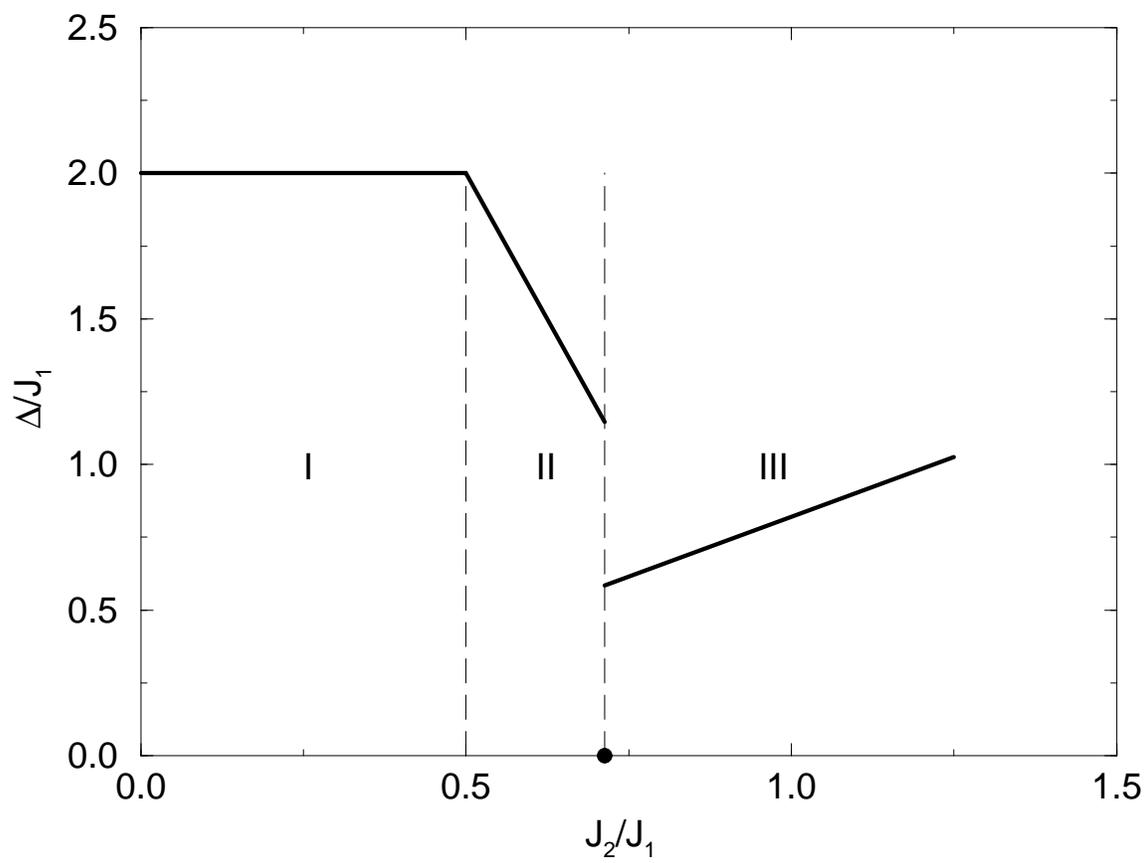
\end{document}